\documentclass[aps,prb,twocolumn,showpacs,amsmath,floatfix]{revtex4}
\usepackage{amsmath,graphics,epsfig,color,verbatim,ulem}
\usepackage{dcolumn,multirow}
\usepackage{subfigure}

\begin{document}

\title{Large spin-phonon coupling and magnetically-induced phonon anisotropy 
in Sr$M$O$_3$ perovskites ($M$=V,Cr,Mn,Fe,Co)}

\author{Jun Hee Lee}
\email{jhlee@physics.rutgers.edu}
\author{Karin M. Rabe}
\affiliation{Department of Physics and Astronomy,
Rutgers University, Piscataway, New Jersey 08854-8019, USA}

\marginparwidth 2.7in

\marginparsep 0.5in

\begin{abstract}

First-principles calculations reveal large zone-center spin-phonon coupling and magnetically-driven phonon 
anisotropy in cubic perovskites Sr$M$O$_3$ ($M$=V,Cr,Mn,Fe,Co). 
In particular, the frequency and splitting of the polar Slater mode is found to depend strongly on magnetic ordering. 
The coupling is parameterized in a crystal-structure-dependent Heisenberg model and its main features seen to arise from the Goodenough-Kanamori rules. 
This coupling can be expected to produce distinct low-energy alternative phases, resulting in a rich variety of coupled magnetic, 
structural and electronic phase transitions driven by temperature, stress, electric field and cation substitution. 

\end{abstract}


\maketitle

Multiferroic phases, which exhibit multiple orderings such as spontaneous polarization, magnetization, and strain, are currently 
the subject of much active research \cite{PhysToday, Advances}. 
The main interest lies not in the coexistence of the orderings in the same material but rather, in the couplings between the orderings. In combination with direct
control of the orderings by macroscopic fields and stresses, these couplings lead to physical properties including linear magnetoelectric response, magnetodielectric
response and cross-switching of magnetization by electric fields or polarization by magnetic fields. 
Furthermore, multiferroic systems can exhibit coupled phase
transitions in which the system transforms from the bulk phase to an alternative low-energy phase with distinct ordering of multiple types, with functional behavior
at the phase boundary arising from switching between the two phases.
This behavior offers both fundamental scientific interest and the possibility of novel multifunctionality for technological applications such as sensors, actuators
and information storage.

One major challenge is that magnetostructural coupling is generally quite small. 
Magnetocrystalline anisotropy, which measures the coupling of the direction of the
spin magnetization to the lattice, depends on the spin-orbit interaction and is extremely small 
for all but the highest atomic number elements. Spontaneous
polarization is invariant under time reversal symmetry, the principal symmetry broken 
by spontaneous magnetization, and conversely, spins are invariant under
inversion, the principal symmetry broken by spontaneous polarization. 
In improper multiferroics such as Tb$_2$MnO$_5$ \cite{ImproperFE}, complex magnetic ordering
breaks inversion symmetry and produces polarization, but the induced polarization is tiny. 
Moreover, the frustration needed to produce this magnetic ordering leads to
low transition temperatures, while higher operating temperatures are needed for practical applications.

On the other hand, certain changes in interatomic distances and bond angles can have a large effect on exchange couplings, leading to 
changes in magnetic ordering energies. 
Systems exhibiting this magnetostructural coupling can be efficiently identified by first-principles calculations\cite{Umesh08,Hena08,ours-SMO,Pablo11} 
of the change in phonon frequency with magnetic order, referred to as spin-phonon coupling. 
Systems where the energy of these structural changes is comparable to the change in magnetic energy 
can have large responses to applied fields and the possibility of coupled phase transitions.
For example, in EuTiO$_3$ \cite{Rabe06} and SrMnO$_3$ \cite{ours-SMO}, it was found that as a result of spin-phonon coupling, a ferromagnetic phase 
with a ferroelectric polar distortion is competitive in energy with the bulk antiferromagnetic paraelectric phase, 
leading to a coupled phase transition driven by epitaxial strain.

In this paper, we report a systematic first-principles study of the spin-phonon coupling of zone-center polar phonons in the cubic perovskites 
Sr$M$O$_3$ ($M$=V,Cr,Mn,Fe,Co), with results for non-magnetic SrTiO$_3$ included for comparison.  
For all magnetic materials studied, the Slater-type phonon mode is found to exhibit the largest spin-phonon coupling, as measured by the frequency shift between the
FM and $G$-AFM ordered phases. Further study of SrMnO$_3$ and SrCoO$_3$, which show the largest shifts, reveals pronounced magnetically-induced phonon anisotropy.
This large spin-phonon coupling is parameterized with a crystal-structure-dependent Heisenberg model 
and analyzed with the help of the Goodenough-Kanamori rules \cite{GK}. 
Experimentally observable consequences of the spin-phonon coupling are discussed, in particular the possibility of low-energy alternative phases and coupled phase
transitions as a function of applied fields, stress, and cation substitution.

\renewcommand{\arraystretch}{1.2}
\begin{table}
\caption{Cubic perovskite lattice constant $a_0$ (in \AA), magnetic ordering type and band gap (in eV), 
compared with available experimental data (in parentheses). 
The lattice constant was calculated for the computed magnetic type given in the table, except for that of SrCrO$_3$, which was computed 
for $G$-type ordering and compared with measurement in the paramagnetic phase. 
If the system is metallic, the gap is reported as $m$. The relative energies $\Delta E$ of alternative magnetic orderings specified 
by the letter in parentheses (in meV/f.u.) are computed at the given computed lattice constant. 
The experimentally determined ordering temperatures $T_c$ (in K) are included. 
Experimental data for SrVO$_3$ is from Refs.~\onlinecite{SV-mag,SV-latt}, for SrCrO$_3$ from Ref.~\onlinecite{SCr-mag}, 
for SrMnO$_3$ from Refs.~\onlinecite{SMn-latt,SMn-mag}, for SrFeO$_3$ from Refs.~\onlinecite{SFe-gap,SFe-latt}, 
for SrCoO$_3$ from Ref.~\onlinecite{SCo-latt}, and for SrTiO$_3$ from Refs.~\onlinecite{STO-latt,STO-gap}.} 

\begin{ruledtabular}
\begin{tabular}{lllccccccccccccc}
&           & $\;\;\;\;\;$$a_0$ & order     & gap      &  $\Delta E$  &$T_{\rm c}$\\ 
\hline 
& SrVO$_3$  &   3.88(3.85)      &  F(SDW)   & $m(m)$   &  37(G)       & 85~K      \\ 
& SrCrO$_3$ &   3.86(3.82)      &  C(C)     & $m(m)$   &  7(G),53(F)  & 40~K      \\
& SrMnO$_3$ &   3.85(3.81)      &  G(G)     & 0.45($i$) &  86(F)       & 233~K     \\
& SrFeO$_3$ &   3.89(3.85)      &  F(hel)   & $m(m)$   & 220(G)       & 134~K     \\
& SrCoO$_3$ &   3.84(3.83)      &  F(F)     & $m$      & 258(G)       & 305~K     \\
& SrTiO$_3$ &   3.95(3.91)      &  N(N)     & 1.8(3.3) &   -          &   -       \\             
\end{tabular}
\end{ruledtabular}
\label{information}
\end{table}

\renewcommand{\arraystretch}{1.0}
First-principles calculations were performed using density-functional 
theory within the
generalized gradient approximation GGA+$U$ method \cite{GGAU}
with the Perdew-Becke-Erzenhof parameterization \cite{PBE}
as implemented in
the $Vienna$ $Ab$ $Initio$ $Simulation$ $Package$ 
(VASP-4.6)~\cite{Kresse2,Kresse3}.  
We use the Dudarev~\cite{Dudarev} implementation
with on-site Coulomb interaction $U$=2.5 eV
and on-site exchange interaction $J_H$=1.0 eV
to treat the localized 3$d$ electron states. 
The value of $U$ for Mn was adjusted to 2.7 eV to fit the experimental magnetic moment of 2.6$\pm$0.2 $\mu_B$ for SrMnO$_3$ \cite{SMn-mag}. 
For Ti, we use $U$ = 0.
The projector augmented wave (PAW) potentials \cite{Kresse1} explicitly
include 10 valence electrons for Sr (4$s^2$4$p^6$5$s^2$), 6 for oxygen (2$s^2$2$p^4$), 
13 for V (3$s^2$3$p^6$3$d^4$4$s^1$), 12 for Cr (3$p^6$3$d^5$4$s^1$),
13 for Mn (3$p^6$3$d^5$4$s^2$), 14 for Fe (3$p^6$3$d^7$4$s^1$), 
and 9 for Co (3$d^8$4$s^1$). 
The zone-center phonon frequencies of the ideal cubic perovskite 
reference structures with various magnetic orderings 
were computed using the frozen phonon method in supercells with $\sqrt 2 \times \sqrt 2 \times \sqrt 2$, $\sqrt 2 \times \sqrt 2 \times 1$, 
$1 \times 1 \times 2$, and $1 \times 1 \times 1$ primitive perovskite cells for $G$-AFM, $C$-AFM, $A$-AFM, and FM, respectively. 
Spin-orbit coupling was not included, so that the phonon wavevectors for the primitive perovskite cell remain good quantum numbers. 
We obtained the modes at $\vec q$ = 0 by uniformly displacing all atoms in related by translation symmetries of the primitive perovskite structure. 
For each supercell, Monkhorst-Pack (M-P) $k$-point meshes \cite{kpt} were chosen to obtain phonon frequencies 
converged to within a few cm$^{-1}$. 

The computed ground state cubic lattice parameters and magnetic orderings are reported in Table~\ref{information} and compared with experimental information. 
The lowest energy magnetic ordering is correctly computed in all cases except SrVO$_3$\cite{SV-mag} and SrFeO$_3$\cite{SFe-latt}, 
which are observed to have ferromagnetic spin density wave and complex helical ordering, respectively, not included in the first-principles analysis. 
The computed relative energies of higher-energy magnetic orderings are reported and are seen to correlate roughly with 
the experimentally-observed magnetic-ordering temperatures given in Table~\ref{information}.
The computed lattice constants are larger than experimental values by about 1\%, as typical for GGA calculations of oxides.
Consistent with experiment, all the magnetic compounds are found to be metallic 
except for SrMnO$_3$, which has a computed gap of 0.45 eV, comparable to that found in a previous study\cite{SMO-gap}.

\renewcommand{\arraystretch}{1.3}
\begin{table}[t]
\caption{ Computed IR-active polar phonon frequencies (cm$^{-1}$) of $G$-AFM and FM states in cubic SrMO$_3$ perovskites. 
The calculations were done with computed lattice constants given in Table~\ref{information}.} 
\begin{tabular}{lccccc}
\hline                                             
\hline                                             
&$\;\;\;$SrTiO$_3$ ($d^0$)$\;\;\;$&\multicolumn{2}{c}{$\;\;\;$SrVO$_3$ ($d^1$)$\;\;\;$}&\multicolumn{2}{c}{$\;\;\;$SrCrO$_3$ ($d^2$)}\\                
&$\;\;\;$ non-mag.$\;\;\;$ &$\;\;\;$ $G$-AF & FM$\;\;\;$ & $\;\;\;$$G$-AF & FM \\
\hline                                 
Slater &$\;\;\;$130 $i$$\;\;\;$& $\;\;\;$329 & 290 $\;\;\;$& $\;\;\;$287 & 187  \\                
Last   &$\;\;\;$  144  $\;\;\;$& $\;\;\;$148 & 147 $\;\;\;$& $\;\;\;$156 & 145  \\                
Axe    &$\;\;\;$  508  $\;\;\;$& $\;\;\;$541 & 546 $\;\;\;$& $\;\;\;$525 & 512  \\                
\end{tabular}
\begin{tabular}{lcccccc}
\hline                                             
\hline                                             
&\multicolumn{2}{c}{$\;\;\;$SrMnO$_3$ ($d^3$)$\;\;\;$}&\multicolumn{2}{c}{$\;\;\;$SrFeO$_3$ ($d^4$)$\;\;\;$}&\multicolumn{2}{c}{$\;\;\;$SrCoO$_3$ ($d^5$)}\\                
 & $\;\;\;$$G$-AF & FM$\;\;\;$ & $\;\;\;$$G$-AF & FM $\;\;\;$& $\;\;\;$$G$-AF & FM \\
\hline                                             
Slater & $\;\;\;$120 &122 $i$ $\;\;\;$& $\;\;\;$248 & 231 $\;\;\;$ &$\;\;\;$176 $i$ & 203 \\                
Last   & $\;\;\;$165 & 165    $\;\;\;$& $\;\;\;$148 & 150 $\;\;\;$ &$\;\;\;$ 155    & 154 \\                
Axe    & $\;\;\;$475 & 486    $\;\;\;$& $\;\;\;$506 & 505 $\;\;\;$ &$\;\;\;$ 508    & 498 \\                
\hline                                             
\hline                                             
\end{tabular}
\label{frequency}
\end{table}

\renewcommand{\arraystretch}{1.0}
In Table~\ref{frequency} we report the computed zone-center polar phonon frequencies for FM and $G$-AFM ordering. 
The phonons are labeled 
according to the character of the eigenmode. 
The Slater mode \cite{Slater} involves the oscillation of the $B$-cation against the oxygen-octahedron 
network. This is the soft mode which is closely associated with the ferroelectric instability in perovskites such as BaTiO$_3$ and KNbO$_3$.   
The Last mode \cite{Last} involves the oscillation of the $A$-cation against the oxygen-octahedron network. 
It is this mode which is the soft mode for $A$-site ferroelectricity in perovskites such as PbTiO$_3$ or BiFeO$_3$. 
The Axe mode \cite{Axe}, at the highest frequency, corresponds to distortion of oxygen octahedra. 

While the Last and the Axe mode frequencies are, as would be expected from their eigenmode character, relatively insensitive to 
the $B$-site $d$-occupancy and magnetic ordering, the Slater mode shows a dramatic shift with change in $d$-occupation 
from compound to compound, and with change in magnetic ordering for a given compound. 
The Slater modes are highest in frequency for $M$=V ($d^1$), $M$=Cr ($d^2$) and $M$=Fe ($d^4$); 
with cubic symmetry $t_{2g}$-$e_{g}$ splitting, these cases involve partially occupied sublevels. 
For $M$=Mn ($d^3$) and $M$=Co ($d^5$), the Slater modes are substantially lower in frequency. 
The frequency shifts of the Slater mode between FM and $G$-AFM ordering are more than 100 cm$^{-1}$ for SrCrO$_3$, SrMnO$_3$ and SrCoO$_3$. 
In SrVO$_3$ and SrFeO$_3$, though the shift is smaller than for the other compounds, it is still significantly larger than for the other two modes.
The low frequencies and the large magnetic ordering shifts in SrMnO$_3$ and SrCoO$_3$ lead, in fact, to ferroelectric instabilities 
in the FM and $G$-AFM orderings, respectively. However, in each case these are not the lowest energy magnetic orderings, 
and thus the ground state structure does not exhibit a polar distortion.
\begin{figure}
\begin{center}
\includegraphics[width=7.7cm,trim=0mm 8mm 0mm 0mm]{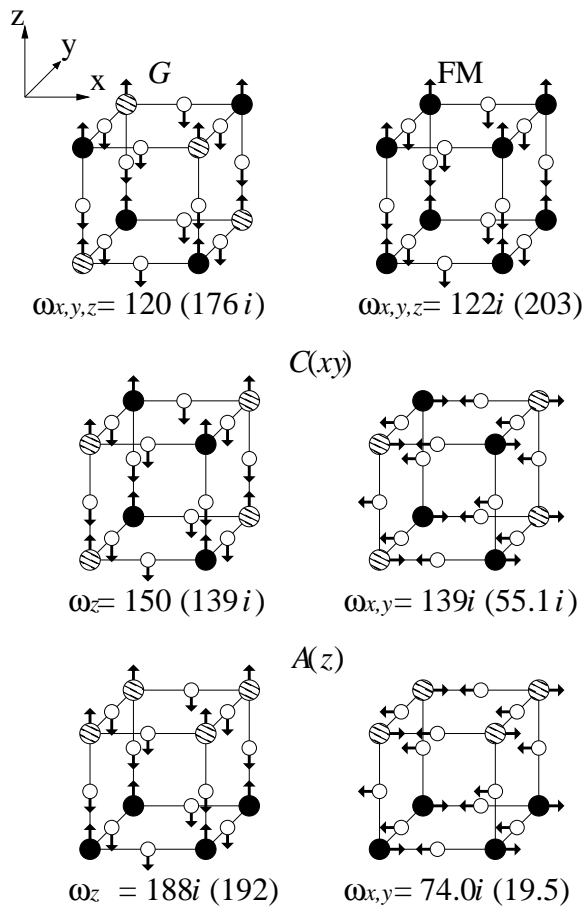}
\end{center}
\caption{(Color online) Computed frequencies for the Slater mode(s) for various magnetic orderings of cubic SrMnO$_3$, with results for SrCoO$_3$ given in parentheses. Arrows represent atomic displacements.   
Black and shaded circles represent up and down spins on the magnetic $B$-site in $AB$O$_3$ perovskites; small open circles represent oxygen atoms.
Frequencies are given in cm$^{-1}$.} 
\label{coupling}
\end{figure}
Next, we computed the polar phonon frequencies with two lower-symmetry magnetic orderings for SrMnO$_3$ and SrCoO$_3$, the two systems with the largest frequency
shift with magnetic ordering.
Specifically, we considered $C$-type, with AFM ordering in the $xy$ plane and FM ordering along the normal direction, giving periodicity $\vec{q}=(\pi/a,\pi/a,0)$,
and $A$-type, with FM ordered planes ordered antiferromagnetically along the normal direction, giving periodicity $\vec{q}=(0,0,\pi/a)$.

The magnetic ordering and the frequencies of the associated Slater modes are shown in Figure~\ref{coupling}. 
The symmetry lowering due to the $C$- and $A$-AFM orderings lifts the degeneracy of the Slater mode in each case by a considerable amount. 
This frequency splitting, which occurs even without any symmetry-lowering relaxation of the crystal structure, is referred to 
as magnetically-induced phonon anisotropy \cite{Massidda,Craig06}. 
In $C(xy)$-AFM SrMnO$_3$, the Slater mode with polarization along $z$ is much stiffer than that with polarization in the plane, 
which is unstable; the splitting is reversed in SrCoO$_3$, with both modes unstable. 
In contrast, in $A(z)$-AFM SrMnO$_3$ both modes are unstable though the ferroelectric instability along the $z$-direction is favored, 
while in SrCoO$_3$ the Slater mode with polarization along $z$ is much stiffer than that with polarization in the plane, which has a very low frequency. 

Each magnetic ordering and atomic displacement pattern can be characterized by the $B$ spins and atomic displacements of $B$ and oxygen  
relative to the line $B$-O-$B$: same spin and transverse displacement (FT), opposite spin and transverse displacement (AT), 
same spin and longitudinal displacement (FL) and opposite spin and longitudinal displacement (AL).
For SrMnO$_3$, low frequency is primarily associated with the FT bonds, and to a lesser extent the AL bonds, 
which predominate in $A(z)$ magnetic ordering with displacements along the $z$ direction. 
On the other hand, for SrCoO$_3$ low frequency is associated with the AT bonds. 

For a precise quantitative description of the spin-phonon coupling, we use the model parameterization of Ref.~\onlinecite{Craig06}. 
The magnetic interactions are described by a Heisenberg model for the spins on the $B$ site\cite{mean-field,Heisen}, 
with exchange couplings determined by the computed energy differences between selected magnetic orderings. 
The coupling to the lattice originates from the dependence of the exchange coupling parameters on the atomic positions. 
Thus, changes in atomic displacement patterns result in changes in magnetic ordering energies; conversely, changes 
in magnetic orderings result in changes in the energetics of atomic displacements, in particular the interatomic force constants, 
producing the spin-phonon coupling.

For the cubic magnetic perovskite compounds considered here, the model energy is written
$E=E_0+E_{ph}+E_{spin}$. 
\noindent 
$E_0$ is the energy of the paramagnetic cubic perovskite structure.
The phonon term $E_{ph}=\frac{1}{2} \sum_{\tau \alpha \tau' \alpha'} K_{\tau \alpha \tau' \alpha'} u_{\tau \alpha}u_{\tau' \alpha'}$ is 
written here to include only primitive-perovskite cell zone-center modes, so that $\tau$ indexes the five atoms in the unit cell and $\alpha$ indicates the Cartesian
direction of the sublattice displacement $\vec u_\tau$.
$E_{spin} = -\sum_{i\neq j} J_{ij} \vec{S_i} \cdot \vec{S_j}$ ($S$=3/2), where $i$ and $j$ index 
the $B$ site spins, is the Heisenberg model for the magnetic ordering energy. 
The exchange couplings are included up to third neighbors to reproduce the relative magnetic ordering energies 
in the cubic reference structure $\vec u_\tau$=0. 
For SrMnO$_3$, this yields $J_1$=-3.1 meV/f.u., $J_2$=-0.16 meV/f.u. and $J_3$=0.0027 meV/f.u.
For SrCoO$_3$ $J_1$=8.8 meV/f.u., $J_2$=0.59 meV/f.u. and $J_3$=0.69 meV/f.u.
The decay with interspin spacing is considerably less rapid in SrCoO$_3$ than in SrMnO$_3$, 
as expected for a metallic system. In the following, we will consider the atomic-displacement dependence
only for the first two neighbor interactions. 

\renewcommand{\arraystretch}{1.2}
\begin{table*}[!t]
\caption{$J''_{1x}$ (upper) and $J''_{1z}$ (lower) of SrMnO$_3$ (left) and SrCoO$_3$ (right). Unit is in 10$^{-4}$ eV/\AA$^2$.}

  \subtable
     {
      \begin{tabular}{lccccc}
      \hline
      \hline
        $J''_{1x}$ $\;\;\;\;$ & Sr $\;\;\;\;$  & Mn $\;\;\;\;$   & O$_x$ $\;\;\;\;$& O$_y$ $\;\;\;\;$& O$_z$  \\
      \hline
       Sr    $\;\;\;\;$       & -61 $\;\;\;\;$ &  -14 $\;\;\;\;$ & 17   $\;\;\;\;$ & 70  $\;\;\;\;$ & -9       \\
       Mn    $\;\;\;\;$       & -14 $\;\;\;\;$ & 2400 $\;\;\;\;$ & -2000$\;\;\;\;$ & 310 $\;\;\;\;$ & -650     \\
       O$_x$ $\;\;\;\;$       &  17 $\;\;\;\;$ & -2000$\;\;\;\;$ & 1500 $\;\;\;\;$ & -43 $\;\;\;\;$ &  530     \\
       O$_y$ $\;\;\;\;$       &  70 $\;\;\;\;$ & 310  $\;\;\;\;$ &  -43 $\;\;\;\;$ & -270$\;\;\;\;$ &  -80     \\
       O$_z$ $\;\;\;\;$       & -9  $\;\;\;\;$ & -650 $\;\;\;\;$ & 530  $\;\;\;\;$ & -80 $\;\;\;\;$ &  210     \\
      \hline
      \end{tabular}
     }
   \subtable
      {
      \begin{tabular}{lccccc}
      \hline
      \hline
       $J''_{1x}$ $\;\;\;\;$    & Sr  $\;\;\;\;$  & Co $\;\;\;\;$   & O$_x$  $\;\;\;\;$  & O$_y$ $\;\;\;\;$& O$_z$ \\
      \hline
        Sr   $\;\;\;\;$ & -7 $\;\;\;\;$   & -40 $\;\;\;\;$  &  66  $\;\;\;\;$    & -17 $\;\;\;\;$  & -5    \\
        Co   $\;\;\;\;$ & -40$\;\;\;\;$   &-3300$\;\;\;\;$  &  2400$\;\;\;\;$    & -170$\;\;\;\;$  & 1100  \\
        O$_x$$\;\;\;\;$ &  66$\;\;\;\;$   & 2400$\;\;\;\;$  & -1700$\;\;\;\;$    & 49  $\;\;\;\;$  & -820  \\
        O$_y$$\;\;\;\;$ & -17$\;\;\;\;$   & -170$\;\;\;\;$  &  49  $\;\;\;\;$    & 83  $\;\;\;\;$  &  50   \\
        O$_z$$\;\;\;\;$ & -5 $\;\;\;\;$   & 1100$\;\;\;\;$  & -820 $\;\;\;\;$    & 50  $\;\;\;\;$  & -330  \\
      \hline
      \end{tabular}
      }

    \subtable
      {
      \begin{tabular}{lccccc}
      \hline
      \hline
       $J''_{1z}$$$ $\;\;\;\;$  & Sr $\;\;\;\;$  & Mn $\;\;\;\;$   & O$_x$ $\;\;\;\;$& O$_y$ $\;\;\;\;$& O$_z$  \\
      \hline                                                
       Sr     $\;\;\;\;$      & -30 $\;\;\;\;$ &  -75 $\;\;\;\;$ & 26  $\;\;\;\;$  & 26  $\;\;\;\;$ &  54     \\
       Mn     $\;\;\;\;$      & -75 $\;\;\;\;$ & -1500$\;\;\;\;$ & 220 $\;\;\;\;$  & 220 $\;\;\;\;$ & 1200    \\
       O$_x$  $\;\;\;\;$      &  26 $\;\;\;\;$ &  220 $\;\;\;\;$ & -150$\;\;\;\;$  & -8  $\;\;\;\;$ &  -80    \\
       O$_y$  $\;\;\;\;$      &  26 $\;\;\;\;$ & 220  $\;\;\;\;$ &  -8 $\;\;\;\;$  & -150$\;\;\;\;$ &  -80    \\
       O$_z$  $\;\;\;\;$      &  54 $\;\;\;\;$ & 1200 $\;\;\;\;$ & -80 $\;\;\;\;$  & -80 $\;\;\;\;$ & -1100   \\
      \hline
      \end{tabular}
      }
     \subtable
     {
      \begin{tabular}{lccccc}
      \hline
      \hline
       $J''_{1z}$ $\;\;\;\;$   & Sr $\;\;\;\;$   & Co $\;\;\;\;$   & O$_x$ $\;\;\;\;$   & O$_y$ $\;\;\;\;$& O$_z$  \\
      \hline                                                   
      Sr    $\;\;\;\;$ & 27  $\;\;\;\;$  &  46  $\;\;\;\;$ & -11 $\;\;\;\;$     & -11  $\;\;\;\;$ & -53    \\
      Co    $\;\;\;\;$ & 46  $\;\;\;\;$  &-1100 $\;\;\;\;$ &  250$\;\;\;\;$     &  250 $\;\;\;\;$ &  540   \\
      O$_x$ $\;\;\;\;$ & -11 $\;\;\;\;$  &  250 $\;\;\;\;$ &  150$\;\;\;\;$     & 19   $\;\;\;\;$ & -410   \\
      O$_y$ $\;\;\;\;$ & -11 $\;\;\;\;$  &  250 $\;\;\;\;$ &  19 $\;\;\;\;$     & 150  $\;\;\;\;$ & -410   \\
      O$_z$ $\;\;\;\;$ & -53 $\;\;\;\;$  &  540 $\;\;\;\;$ & -410$\;\;\;\;$     & -410 $\;\;\;\;$ &  330   \\
      \hline
      \end{tabular}
      }

\label{J-SM-SC}
\end{table*}
\renewcommand{\arraystretch}{1.0}
To investigate the spin-phonon coupling, we expand the model energy in the sublattice displacements $\vec u_{\tau}$. For spin configurations of sufficiently high
symmetry, including the five configurations considered here, the first order terms vanish.
At second order, the full force constant matrix $\tilde{K}=\partial^2 E / \partial u_{\tau \alpha} \partial u_{\tau' \alpha'}$ is given by
\begin{displaymath}
 \tilde{K}_{\tau \alpha \tau' \alpha'} = K_{\tau \alpha \tau' \alpha'}-\sum_{i\neq j}
   J_{ij,\tau \alpha \tau' \alpha'}''u_{\tau \alpha} u_{\tau' \alpha'} <\vec{S_i} \cdot \vec{S_j}>
\end{displaymath}
with $J_{ij\tau \alpha,\tau' \alpha'}'' \equiv \frac{\partial^2 J_{ij}}{\partial u_{\tau \alpha} \partial u_{\tau' \alpha'}}$.
By symmetry, the $15\times 15$ matrices $J_{ij}''$ are block diagonal, with nonzero elements only for sublattice displacements $u_{\tau}$ and $u_{\tau'}$ 
along the same Cartesian direction. Furthermore, cubic symmetry relates the $5\times 5$ block for displacements for one choice of $ij$ and Cartesian displacement direction to the
blocks for other choices of $ij$ and Cartesian displacement direction, so that there are only four independent blocks: two for first neighbor exchange couplings 
(z-displacement blocks $J''_{1x}$ and $J''_{1z}$, with $J''_{1y}$ related to $J''_{1x}$,) and for next nearest neighbors 
($J''_{2\parallel}$ for displacements along $z$ and exchange couplings within the $xy$ plane, and $J''_{2\perp}$ for displacements along $z$ and exchange couplings to spins at greater or lower $z$).
For the spin configurations considered, the force constant matrices have the same block diagonal form as the $J''$ matrices and can be compactly expressed in terms of
$J''$ z-displacement blocks as follows:
\begin{displaymath}
\left[\begin{array}{c} \tilde{K}^{\rm F}(\hat{z}) \\ \tilde{K}^{G}(\hat{z}) \\ \tilde{K}^{C}(\hat{z}) \\ \tilde{K}^{C}(\hat{x}) \\ 
\tilde{K}^{A}(\hat{z}) \\ \tilde{K}^{A}(\hat{y}) \end{array}\right]=
\begin{bmatrix} 1 & -9/2 & -9/2 & -9/2  & -9 & -18  \\
                1 & +9/2 & +9/2 & +9/2  & -9 & -18  \\
                1 & +9/2 & +9/2 & -9/2  & -9 & +18  \\
                1 & +9/2 & -9/2 & +9/2  & +9 &   0  \\
                1 & -9/2 & -9/2 & +9/2  & -9 & +18  \\
                1 & -9/2 & +9/2 & -9/2  & +9 &   0  \\
\end{bmatrix}
\left[\begin{array}{c} K \\ J''_{1x} \\ J''_{1y} \\ J''_{1z} \\ J''_{2\parallel} \\ J''_{2\perp} \end{array} \right]
\end{displaymath}
where $\tilde{K}^{\rm F}$, $\tilde{K}^{G}$, $\tilde{K}^{C(xy)}$,
$\tilde{K}^{A(z)}$ are 5$\times$5 blocks of the full force constant matrices. Using first-principles computations for these matrices, we can determine the $J''$
blocks and $K$, the force constant matrix for paramagnetic ordering $<\vec{S_i} \cdot \vec{S_j}>$=0.

Table~\ref{J-SM-SC} shows 
$J''_{1x}$ and $J''_{1z}$ 5$\times$5 $z$-displacement blocks for SrMnO$_3$ and SrCoO$_3$.
O$_x$ and O$_y$ denote the equatorial oxygens 
in the line $B$-O-$B$ along $x$ and the equatorial oxygen in the line $B$-O-$B$ along $y$, respectively.
From these matrices, we find that the large spin-phonon coupling noted for SrMnO$_3$ and SrCoO$_3$ 
can be mainly attributed to the nearest neighbor exchange interaction 
submatrix $J''_{1x}$, specifically the transverse displacement of the $B$ cations and oxygen atoms in the $B$-O-$B$ bond along $x$.
For a more detailed discussion, we diagonalized the $J''_{1x}$ matrices.
Eigenvalues of the $J''_{1x}$ block
for SrMnO$_3$ are (0.42,-0.04,-0.01,0.005,0.000) with eigenvector (0.00,0.76,-0.62,0.06,-0.21) for the largest eigenvalue.
A phonon with this displacement pattern thus would exhibit the maximum spin-phonon coupling, explaining the large effect for the Slater mode.
The fact that the four of the five eigenvalues are close to zero allows us to approximate the atomic-displacement-dependent part of $J_{1x}$ using this single
eigenvector:
\begin{displaymath}
\Delta J_{1x} \sim 0.066 \big[ (u_{\rm Mn}-u_{{\rm O}_x})
+\frac{1}{3}(u_{\rm Mn}-u_{{\rm O}_z}) \big]^2.
\end{displaymath}
The bond angle reduction represented by the first term 
makes the main contribution to the spin-phonon coupling, with the bond length change represented by the second term being a secondary contribution. 

For SrCoO$_3$, the analysis is very similar.
Eigenvalues of the $J''_{1x}$
of SrCoO$_3$ are (-0.54,0.013,0.005,-0.003,0.000) with eigenvector (0.01,0.78,-0.56,0.03,-0.26) for the largest eigenvalue.
Again, the other eigenvalues are close to zero, allowing the approximation
\begin{displaymath}
\Delta J_{1x} \sim -0.093 \big[ (u_{\rm Co}-u_{{\rm O}_x})
+\frac{1}{3}(u_{\rm Co}-u_{{\rm O}_z}) \big]^2.
\end{displaymath}
As for SrMnO$_3$, the two terms represent the bond-angle reduction and bond-length change contributions, which for SrCoO$_3$ are negative rather than positive.

The importance of the two terms in SrMnO$_3$ can be qualitatively understood 
using the Goodenough-Kanamori rules \cite{GK}. 
The magnetic exchange coupling $J$ is given by
$J \sim t^2/(\Delta -J_H)$ 
where $t$ is the hopping integral, which is proportional to 
the effective wavefunction overlap between the two cations through their shared oxygen, $\Delta$ is the energy difference 
between oxygen and $B$-cation orbitals and $J_H$ is the Hund coupling. 
Reduction of the bond angle (($\Delta \theta)^2$ $\sim$ ($u_{\rm B}-u_{{\rm O}_x}$)$^2$) 
decreases the hopping integral $t$, which reduces the 
magnitude of the exchange coupling $J$. 
Also, displacement of the apical oxygens O$_z$ towards the $B$ cation  
should shift the $B$-cation and O$_z$ orbital energy level in opposite directions 
due to electrostatics; the resulting decrease in the magnitude of $J$ 
arising from the change in $\Delta$. 
For SrCoO$_3$, ferromagnetism can be explained by analogy to the Zener double exchange (DE) mechanism \cite{DE1,DE2} 
with high-spin Co $d^6$ ions and ligand holes of 1/3 per oxygen antiferromagnetically coupled to the Co spin, 
with the mobile holes favoring ferromagnetic alignment of the cobalt spins \cite{inter}. 
If the $d-p$ matrix element is decreased by a reduction in bond angle, or the change-transfer gap is increased 
by displacement of the apical oxygen, this will decrease the magnitude of the magnetic exchange coupling \cite{Zhao2000}. 

While, as discussed above, the Slater mode maximizes the spin-phonon coupling, the other polar modes show very little change with magnetic ordering. \cite{ours-SMO,Kim96,Kamba10} 
The Axe mode consists mainly of
O$_z$ displacements, which reduce the $B$-O$_z$ bond length without changing the $B$-O$_x$-$B$ bond angles. 
The first term does not contribute and the change in $J$ is much
smaller, corresponding to a change only of $\Delta$. 
The Last mode, which is
a displacement of $A$-site cation with respect to the $B$-O$_6$ network, does not change the relevant bond angles or bond lengths, and both terms are negligible. 

The large spin phonon coupling for the Slater mode found from first principles can be experimentally observed in bulk samples by changing the magnetic ordering.
In the paramagnetic phase, application of a magnetic field slightly increases alignment of the spins, leading to a phonon frequency shift 
that is rather small. 
For example, a mean field theory calculation for SrMnO$_3$ gives a shift of approximately 0.1 cm$^{-1}$ with a field of 5 T at $T_{\rm N}$. 
A much more dramatic shift can be observed by changing the temperature through the magnetic phase transition. As the magnetic ordering changes from paramagnetic above
the magnetic ordering temperature to the low-temperature ordered phase, the change in the spin correlations produces a substantial shift in phonon frequency that is
roughly half the $G$-AFM-FM splitting.
Indeed, motivated by these first principles results, far IR measurements have been made on SrMnO$_3$ \cite{Kamba10}, showing the expected hardening of the Slater mode
on cooling through $T_{\rm N}$=250 K.

Another readily measurable physical property is the magnetodielectric effect. For example, in La$_2$MnNiO$_6$, an applied magnetic field shifts the critical temperature of a
phase transition at which the dielectric constant jumps by about 10\%, producing a large magnetodielectric response between the two temperatures.\cite{Rogado05} In addition,
magnetically-induced anisotropy could be directly observed by measuring phonon splittings in a material that has a $C$-AFM bulk phase, for example SrCrO$_3$. In this
material, a large coupling between the magnetic moment and a breathing phonon in a hypothetical $G$-AFM phase has been noted from first principles \cite{Pickett09}. 

In thin films and superlattices, the large spin-phonon coupling can produce a coupled magnetic-ferroelectric phase transition with epitaxial strain as
polarization-strain coupling increases the instability of the polar mode for non-ground-state magnetic ordering (it should be noted that other perturbations, such as
compositional substitution, could also be effective). An epitaxial-strain-induced multiferroic (ferromagnetic-ferroelectric) phase based on this mechanism was
identified from first principles investigation in EuTiO$_3$\cite{Rabe06} and subsequently experimentally confirmed\cite{EuTiO3expt}. 
First-principles investigation of the epitaxial-strain phase diagram of SrMnO$_3$ has shown an analogous multiferroic phase, 
with higher ordering temperature\cite{ours-SMO}. 
In SrCoO$_3$, with a ferromagnetic metallic bulk state, the polarization-strain coupling increases the ferroelectric instability of the insulating antiferromagnetic
phase, driving a coupled magnetic-ferroelectric metal-insulator transition with epitaxial strain, which has been the subject of a first-principles
study\cite{ours-SCO}. 
These systems are expected to exhibit desirable functional behavior at the phase boundary associated with switching between the two phases by applied fields or
stresses. 

In summary, we have presented a first-principles
investigation of the spin-phonon coupling in the Sr $A$-site perovskite compounds Sr$M$O$_3$ ($M$=V,Cr,Mn,Fe,Co). 
The coupling is largest for the Slater phonons, especially for SrMnO$_{3}$ and SrCoO$_{3}$ which show large frequency shifts with changes in magnetic ordering and
magnetically induced phonon anisotropy. We show that this behavior can be qualitatively understood on the basis of Goodenough-Kanamori rules.  
The computed spin-phonon coupling can be experimentally investigated directly through measurement of the phonon frequency shift through the magnetic transition. Other
consequences include the possibility of coupled phase transitions with epitaxial strain or compositional substitution and desirable functional behavior for systems at
the phase boundary. Thus, computation of spin-phonon coupling can be a valuable screening tool in the first-principles design of novel functional and multifunctional
materials.

We would like to thank S.-W. Cheong, C. J. Fennie, D. R. Hamann, A. Stroppa and D. Vanderbilt for valuable discussions. 
This work was supported by MURI ARO Grant W911NF-07-1-0410 and ONR Grant N00014-09-1-0300.

\end{document}